\documentstyle[12pt,epsf]{article}
\begin{document}
\begin{sloppypar}
\begin{center}

{\bf CORRELATION SINGS OF INSTANTONS \\
IN MULTIGLUON PRODUCTION PROCESSES AT HIGH ENERGY\\ }
\vspace{20pt}
{\small V.KUVSHINOV, R.SHULYAKOVSKY      \\

{\it Institute of Physics, Academy of Sciences of Belarus       \\
Minsk 220072 Scorina av.,70\\
e-mail: kuvshino@dragon.bas-net.by \\
e-mail: shul@dragon.bas-net.by }} \\

\vspace{20pt}
\end{center}

General formula for inclusive gluon distribution on rapidities
is obtained for the processes of multigluon production in classical
instanton field with  the first quantum correction.
On the basis of this formula  second correlation function
is calculated in QCD and analysed.
The features of the correlation function behaviour
can be used as a signal of instanton  at HERA.


\vspace{22 pt}
\noindent
{\bf 1.\quad Introduction}
\vspace{10pt}

There exists now large interest to high energy processes induced by
both SU(2)$\times$U(1) weak instantons~[1-2] and SU(3) strong 
instantons~[3-4] because of their important role in HEP.

As it is known 
violation of baryon and lepton numbers conservation law~[5]
due to quantum anomaly~[6] can be induced by SU(2)$\times$U(1) weak
instantons~[7]
which represent tunnelling processes,
associated with the highly degenerated vacuum structure.
Possibility of the baryon and lepton number nonconserevation
at high energy is connected with the problem of the baryon and antibaryon
asymmetry in the observable part of the Universe~[8].
At low energies (when energy of the process is less
than barrier between different degenerated vacuum stages)
cross section $\sigma^I_{tot}\approx 10^{-78}$~[5].
In high energy particle collisions (in multi TeV regime)
the cross section can increase exponentially. It is associated
with multi $W, Z^0$ and $H$-bosons production in the instanton
field~[1].

On the other hand for strong SU(3) instantons in QCD such phenomenon
can exist at hundreds Mev~[2] and be important in deep
inelastic $ep$-scattering
for decreasing Bjorken variable
$X_{Bj}$ and high photon virtuality $Q^2$~[8].
Search of QCD-instantons has started already in $ep$-collisions
at HERA(H1). The processes have some features:
instanton contribution to structure function $F_2(X_{Bj},Q^2)$
rises strongly with decreasing $X_{Bj}$;
$\sigma_{tot}^I$ strongly peaks with  decreasing of $X_{Bj}$;
hadronic band  emission of semi-hard
partons is isotropic in the instanton rest system; current quark jet
and characteristic flavour, strangenes-$K^0$, charm and muon
flow take place~[4,9,10].

Moreover instanton-induced processes manifest new mechanism 
of multiparticle production and can contribute to 
intermittency exponent~[11].

In this paper we study properties of the
second correlation function as  signature of SU(3)
instanton-induced
multi-gluon state for classical instanton with the
first quantum correction in QCD.

It should be noted, for this effect the problem of taking into account
the hadronization exists. Here it can be solved by the use of
the local parton-hadron duality~[12].

\vspace{18 pt}
\noindent
{\bf 2.\quad Exclusive Distribution on Gluon Rapidities
for Instanton- \\
Induced Multiparticle Production Processes with
the First Quantum Correction}
\vspace{10pt}

\noindent
The following instanton solution is used~[7]:
\begin{equation}
A_{\mu}^{a (cl)}(x)=\frac{2\eta_{\mu \nu}^a(x-z)_{\nu}}{g((x-z)^2+\rho^2)},
\end{equation}
\noindent
where $A_{\mu}^a(x)$ are gluon fields;
$\eta_{\mu \nu}^a$ is a 't Hooft symbol~[5]; $\rho$ and $z_{\nu}$
are size and position of the instanton correspondingly;
$g$ is a constant of strong interaction;  greek indexes
$\mu ,\nu$...=1,2,3,4 are the four-vector indexes, latin indexes
$a,b...$=1,2,...,8 are SU(3)-group indexes;
subscript "cl" denotes quasiclassical approximation.

It is convenient to use reduction formula [13] for the
calculation of the amplitude
of n gluon production in instanton field

$$
T_{\ \mu_1\mu_2...\mu_n}^{a_1a_2...a_n}(k_1,k_2,...,k_n)=
\int \prod_{j=1}^ndy_je^{ik_jy_j}(k_j^2+m^2)\times
$$
$$
\times \int [DA]e^{-S[A]}A_{\mu_1}^{a_1}(x)...A_{\mu_n}^{a_n}(x)=
$$
$$
=\int \prod_{j=1}^ndy_je^{ik_jy_j}(k_j^2+m^2) \int [DA]e^{-
S[A]}(A_{\mu_1}^{a_1(cl)}(x)+A_{\mu_1}^{a_1(qu)}(x))...\times
$$
$$
\times(A_{\mu_n}^{a_n(cl)}(x)+A_{\mu_n}^{a_n(qu)}(x))= \int 
\prod_{j=1}^ndy_je^{ik_jy_j}(k_j^2+m^2)\times
$$
$$
\times \int [DA]e^{-S[A]}A_{\mu_1}^{a_1(cl)}(x)...A_{\mu_n}^{a_n(cl)}(x)+\int 
\prod_{j=1}^ndy_je^{ik_jy_j}(k_j^2+m^2)\times
$$
\begin{equation}
\times \int [DA]e^{-
S[A]}A_{\mu_1}^{a_1(qu)}(x)A_{\mu_2}^{a_2(qu)}(x)A_{\mu_3}^{a_3(cl)}(x)...A_{\mu
_n}^{a_n(cl)}(x)+...\quad .
\end{equation}

\noindent
In formula (2) we consider $A_{\mu }^a(x)=A_{\mu }^{a(cl)}(x)+A_{\mu 
}^{a(qu)}(x)$,
where $A_{\mu }^{a(qu)}(x)$ is small fluctuations near classical field $A_{\mu 
}^{a(cl)}(x)$;
subscript "qu" denotes quantum correction~[14].
The first term in formula (2) corresponds to the main (quasiclassical)
approximation and is given by the following expression~[15]:

\begin{equation}
T_{\ 
\mu_1\mu_2...\mu_n}^{(cl)a_1a_2...a_n}(k_1,k_2,...,k_n)=\,(C_1)^n\prod_{j=1}^n
\eta_{\mu_j \nu_j}^{a_j}k_j^{\nu_j},
\end{equation}

\noindent
where $k_1,k_2,...,k_n$ are the 4-momenta of the produced gluons,
$k_j=(\vec k_j, iE_j)$, $C_1=4\pi^2 i\rho^2 /g$,
$m$ is effective gluon mass.

In the formula (3) we as usual do not write $\delta$-function, which is 
connected
with law of conservation of 4-momentum~[1]. For the correct normalizing
we must rewrite formula (3) in the following way:

$$
T_{\ 
\mu_1\mu_2...\mu_n}^{(cl)a_1a_2...a_n}(k_1,k_2,...,k_n)=\,(C_1)^n\prod_{j=1}^n
\eta_{\mu_j \nu_j}^{a_j}k_j^{\nu_j}\Theta (n_{max}-n),
\eqno(3')
$$

\noindent
where $n_{max}=\sqrt{s}/m$.

Then the second term in (2) is the first quantum correction to the
classical amplitude. It is calculated on the basis of gluon
propogation function in instanton field~[16] and is given
by the following formula~[2] (for two produced gluons):
\begin{equation}
T_{\ \mu \nu}^{(qu)ab}(k_1,k_2)=(C_2)^2\eta_{\mu \alpha}^d\eta_{\nu 
\beta}^dk_1^{\alpha}k_2^{\beta}\varepsilon^{abc}\eta_{\kappa 
\lambda}^c\frac{k_1^{\kappa}k_2^{\lambda}}{(k_1,k_2)^2},
\end{equation}
\noindent
where $C_2=2\pi i\rho$; $(k_1,k_2)=(\vec k_1\vec k_2)-E_1E_2$.

We make the following natural assumption (in laboratory subsystem
in $ep$-collision)~[17]:

$$
(k_i^L)^2\gg(\vec k_i^T)^2\gg m^2,\qquad k^T_i\equiv \quad \mid \vec 
k_i^T\mid\,=\,k^T,
$$
\begin{equation}
E_i\approx k^Tchy_i,\qquad k_i^L\approx k^Tshy_i.
\end{equation}
Let us write the expressions for the probabilities of the gluon production
processes going through instanton mechanism in dependence on
rapidity variables:

$$
P_n(y_1,...,y_n)\,=\,\mid T_n^{(cl)}(y_1,...,y_n)\,+\,T_n^{(qu)}(y_1,...,y_n) 
\mid^2
\approx
$$
$$
\approx 
T_n^{(cl)}(y_1,...,y_n)\Bigl[T_n^{(cl)}(y_1,...,y_n)\Bigr]^*+T_n^{(qu)}(y_1,...,
y_n)\Bigl[T_n^{(cl)}(y_1,...,y_n)\Bigr]^*+
$$
\begin{equation}
+ 
T_n^{(cl)}(y_1,...,y_n)\Bigl[T_n^{(qu)}(y_1,...,y_n)\Bigr]^*\,=\,A^n\prod_{j=1}^
nch2y_j\Theta(n_{max}-n)-
\end{equation}
$$
-\alpha A^{n-2}\Bigl[\Pi (y_1,y_2)\,ch2y_3...ch2y_n+...+\Pi (y_{n-
1},y_n)\,ch2y_1...ch2y_{n-2}\Bigr] \times
$$
$$
\times \Theta (n_{max}-n);
$$

\noindent
where $A=3(k^T)^2(C_1)^2$, $\alpha=8(C_1)^2(C_2)^2(k^T)^2$,
$\Pi (y_1,y_2)=sh(y_1-y_2)\times \\
\times th(y_1-y_2)$.
In the formula (6) the first term corresponds to the classical 
approximation and the others are the first quantum correction
contribution.

\vspace{18pt}
\noindent
{\bf 3.\quad Inclusive Distribution and Second Correlation 
Function for Instanton-Induced Multigluon Production
Processes with the First Quantum Correction} 
\vspace{10pt}

In order to calculate inclusive distribution $\rho_n(y_1,...,y_n)$
we must by standart method
take into account all canals of the process.
In our case we obtain:

$$
\rho_n(y_1,...,y_n)=A^nch2y_1...ch2y_n[\Theta (n_{max}-n)+R_0(n)]-
$$
$$
-\alpha A^{n-2}\bigl[\Pi (y_1,y_2)ch2y_3...ch2y_n+...+\Pi (y_{n-
1},y_n)ch2y_1...ch2y_{n-2}\bigr]\times
$$
$$
\times(\Theta (n_{max}-n)+R_0(n))-\frac{\alpha}{2}A^nch2y_1...ch2y_n\Pi 
(Y)R_2(n)-
$$
\begin{equation}
-\alpha A^{n-1}\bigl[Q(Y,y_1)ch2y_2...ch2y_n+...+Q(Y,y_n)ch2y_1...ch2y_{n-
1}\bigr]R_1(n),
\end{equation}

\noindent
where we denoted

$$
\Pi (Y)=\int \limits_{-Y}^Y\Pi (y_1,y_2)=sh(y_1-y_2)th(y_1-y_2)dy_1dy_2=
$$
$$
=4(sh(\frac{Y}{2}))^2-\pi Y+8\sum_{k=0}^{\infty }\frac{(-1)^k}{(2k+1)^2}exp(-
\frac{Y}{2}[2k+1])sh(\frac{Y}{2}[2k+1]);
$$
$$
Q(Y,y)=2shYchy+arcth[sh(y-Y)]-arcth[sh(y+Y)];
$$
$$
R_i(n)=\sum_{m=1}^{\infty}\frac{[Ash2Y]^{m-i}}{(m-i)!}\Theta (n_{max}-n-m), 
$$
\begin{equation}
i=0,1,2, \qquad (0\le R_2(n)\le R_1(n)\le R_0(n)\le 1).
\end{equation}

Two-particle correlation function in dependence on
rapidities

$$
C_2(y_1,y_2)=\rho_2(y_1,y_2)-\rho_1(y_1)\rho_1(y_2)
$$

\noindent
if $n_{max}\gg 1$ and $\Theta (n_{max}-1)=\Theta (n_{max}-2)=1; R_i(1)\approx 
R_i(2)\approx 1;\quad i=0,1,2,$
has the following form:

$$
C_2(y_1,y_2)=[-2A^2+\frac{3}{2}\alpha A^2\Pi (Y)]ch2y_1ch2y_2-
4\alpha A\pi ch(y_1+y_2)th(y_1-y_2)-
$$
$$
-2\alpha sh(y_1-y_2)th(y_1-y_2)+2\alpha AshY[chy_1ch2y_2+chy_2ch2y_1]+
$$
\begin{equation}
+2\alpha 
A\bigl[arctg(\frac{chy_1}{shY})ch2y_2+arctg(\frac{chy_2}{shY})ch2y_1\bigr]                                    
.
\end{equation}

Corresponding curve lies in negative region of the plot
$C_2(y_1,0)$, has maximum at $y_1=0$
and minima at $y_1=+3,5;-3,5$ (see fig.1).Central maximum  corresponds
to the quasiclassical part of (9); two minima are contribution
of the first quantum correction. For the estimation the parameters are taken to 
have
the next values: $\sqrt{s}=50 GeV$, $m=100 MeV$, $\rho =1 GeV^{-1}$, 
$Y \approx 4$, $k^T=0,1 GeV$.

\hspace{2cm}
\begin{minipage}{6cm}
\epsfxsize=3.45in \epsfbox{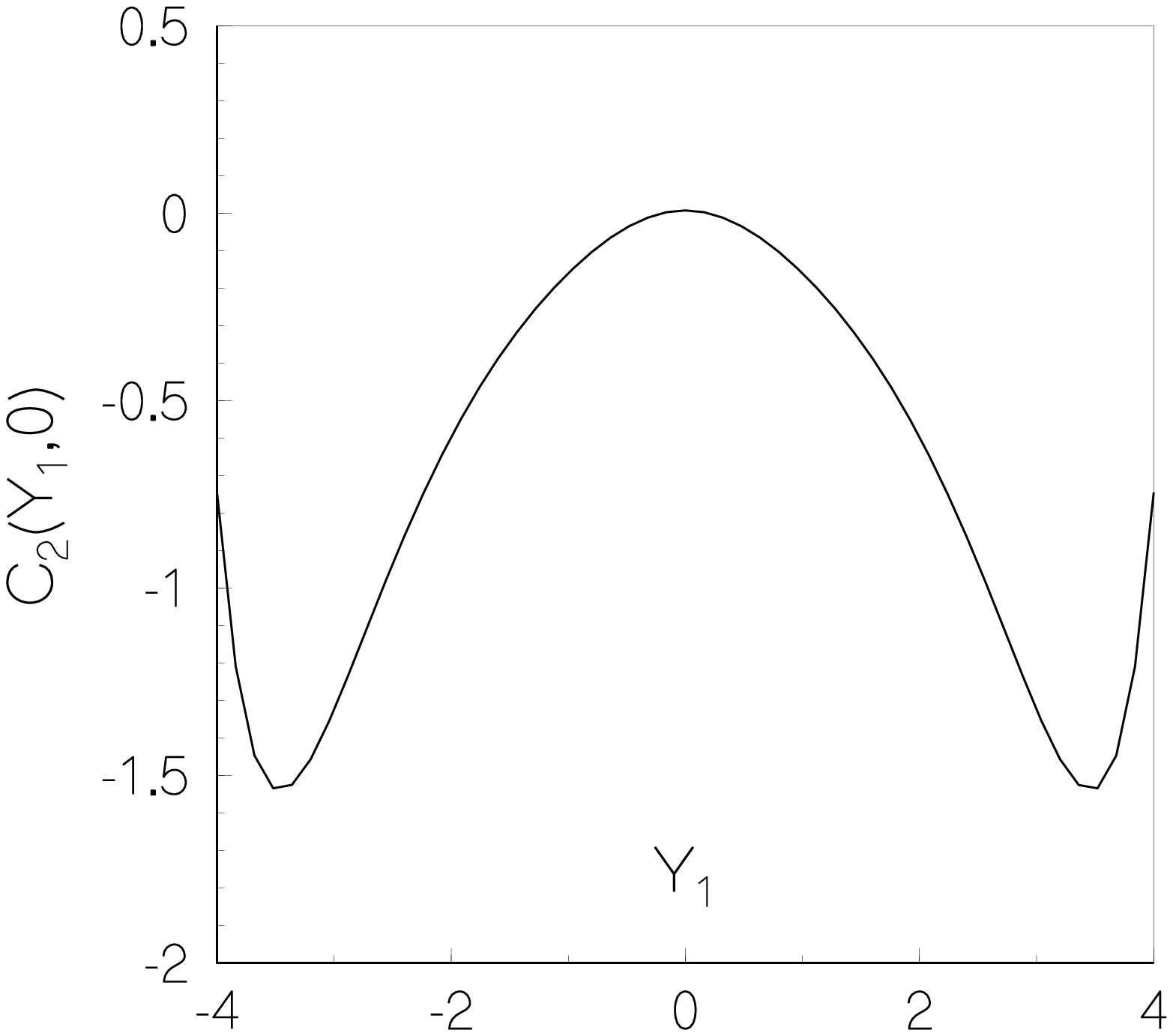}
\end{minipage}

\begin{center}
\vspace*{-1cm}
Fig.1. Correlation function vs. rapidity
$y_1$ of one of the particle, when $y_2=0$.
\end{center}

\vspace{18 pt}
\noindent
{\bf Conclusion}
\vspace{10pt}

\noindent
Thus, in addition to the known footprints of instanton induced
events we have obtained  two gluon correlation function
which is negative and has specific structure at the rest
system of one of the particle. With the help of
local parton-hadron duality the result can be used for
hadrons and is of interest for HERA experiments~[4].

The estimations of
contributions of quantum corrections of higher orders
and of adequate kinematical restrictions, separation
of hadrons from semi-hard quarks and from accompanying
gluons to be used in Monte-Carlo analysis are now in progress.

\vspace{18 pt}

\noindent
The authors are grateful for support in part to International Soros Science 
Education
Program and to Basic Science Foundation of Belarus (Project F95-023).\\

\end{sloppypar}
\end{document}